\documentclass[preprint,superscriptaddress,showpacs]{revtex4}
\usepackage{amssymb}
\usepackage{graphicx}

\newcommand{ \be}{\begin{equation}}
\newcommand{ \ee}{\end{equation}}
\newcommand{\beq}{\begin{eqnarray}}
\newcommand{\eeq}{\end{eqnarray}}
\newcommand{\bem}{\begin{pmatrix}}
\newcommand{\eem}{\end{pmatrix}}
\newcommand{\bmx}{\begin{array}}
\newcommand{\emx}{\end{array}}

\begin{document}

\title{Precursors of order in aggregates of patchy particles}

\author{Oleg A. Vasilyev}
\affiliation{Max-Planck-Institut f{\"u}r Intelligente Systeme, Stuttgart, Germany}
\affiliation{IV. Institut f\"ur Theoretische Physik,  Universit\"at Stuttgart,  Stuttgart, Germany}
\author{Boris A. Klumov}
\affiliation{Joint Institute for High Temperatures, Moscow,  Russia}
\affiliation{Institute for Information Transmission Problems,  Moscow, Russia}
\author{Alexei V. Tkachenko}
\affiliation{Center for Functional Nanomaterials, Brookhaven National Laboratory, Upton, NY,  USA}
\date{\today}

\begin{abstract}
We study computationally the local structure of aggregated systems of patchy particles. By calculating the probability distribution functions of
various rotational invariants  we can identify the precursors of orientation order in amorphous phase. Surprisingly, the
strongest signature of local order is observed for 4-patch particles with tetrahedral symmetry, not for 6-patch particles with the cubic one.
This trend is exactly opposite to their known ability  to crystallize. We relate this anomaly to the observation that  a generic aggregate of
patchy systems has coordination number close to 4. Our results also suggest a significant correlation between rotational order in  the
studied liquids with the corresponding crystalline phases, making this approach potentially useful  for a broader range of patchy systems.
\end{abstract}

\maketitle

The field of colloidal and nanoparticle self-assembly has dramatically
changed over the past decade due to introduction of novel classes of
particles and interactions between them. This includes highly selective
DNA-mediated interactions \cite{DNA1}-\cite{DNA3}, use of wide variety of
particle shapes and so-called "patchy" colloids  \cite{patchy1}-\cite{Patchy
DNA}, as well as combinations of these approaches. The patchy particles have
patterns of chemically distinct regions on their surfaces that results in
directional (covalent-like) interactions. This opens an appealing prospect of
"programming" the symmetry of a desired structure with the symmetry of the
particle. For instance, colloids with tetrahedral arrangement of patches
have been widely studied theoretically \cite{Sciort1}-\cite{glotzer diamond},
as a potential platform for self-assembly of diamond lattice, one of the
best candidates for photonic band gap materials \cite{photonics}. These
studies were in part motivated by experimental demonstration of patchy
colloids with tetrahedral and other symmetries \cite{Patchy2}. Most recently, these
experimental techniques evolved even further due to functionalization of
patches with DNA and resulting selectivity of interactions \cite{Patchy DNA}.

While the equilibrium phase diagrams of such patchy colloids have been
extensively studied computationally, the experimental \ self-assembly of
crystalline morphologies will unavoidably be limited by slow kinetics. By
analogy with self-assembly of colloids and nanoparticles isotropically
functionalized with DNA, one might expect the system to form a random
aggregate initially, and possibly be transformed to a crystal upon
annealing. In this paper we focus on the relatively early stage of self
assembly, and analyze the precursors of crystallinity in a (mostly) random
aggregate of patchy particles. We do this by employing the set of bond order parameters \cite{stein}
which have been successfully
applied to a great variety of physics problems. The parameters are commonly used to determine the degree of
crystallinity of a system, as well as the crystal morphology. In our case,
we apply it to find traces of ordering in liquid phase, which may result in
crystal formation upon annealing. This kind of analysis can potentially
determine the morphology which is preferred kinetically, rather than
energetically. For instance, it is well known that the space of possible
structures for patchy particles with tetrahedral symmetry is highly
degenerate. At least two diamond morphologies, cubic, and hexagonal have nearly the same free energy.
As a result, in majorities of studies (with a rare but noteworthy exception \cite{glotzer diamond}) neither of the
crystals form spontaneously. In such a case, the structure in actual
experiments may be selected kinetically.

Several models have been used to simulate patchy colloids. Most common  are
anisotropic Lennard-Jones potential \cite{doye},\ and Kern-Frenkel model
\cite{KF},\cite{Sciort1},\cite{Sciort2}. In those models,  the patch
geometry can be tuned independently of the interaction range, and the degree
of  directionality is an important parameter of the system. Since the
dependence of the system behavior on the patch size  is not a focus of our
study, here we use a simpler model with point-like patches interacting via
Gaussian potential. The range of the potential automatically determines the
degree of directionality of the bond. Physically, this is a reasonable model
to describe nanoparticles with locally grafted DNA \ molecules, in which
case both patch size and range of interactions is determined by DNA gyration
radius.

{\it Numerical method}~~ Our numerical algorithm realize the description of a
patchy particle as a rigid body.  We represent the motion of a patchy particle of the radius $R$  as a combination of the
displacement of its center and rotation around some axis, passing through its center.
We study two systems with 4 patches (4pch) and 6 patches (6pch).
At the initial time moment orientation locations of four patches
$ {\bf a}_{j}^{(k)}(0)$, $k=1,2,3,4$
for 4pch
particles with tetrahedron symmetry  with respect the center of the $j$-th particle are
${\bf a}^{(1,2)}_{j}(0)=\left( \pm \sqrt{\frac{2}{3}}R,0,\sqrt{\frac{1}{3}}R\right)$,
${\bf a}^{(3,4)}_{j}(0)=\left( 0,\pm \sqrt{\frac{2}{3}}R,-\sqrt{\frac{1}{3}}R\right)$.
For 6pch system with cubical symmetry at initial time moment
patches are located at points
${\bf a}^{(1,2)}_{j}(0)=( \pm R,0,0)$,
 ${\bf a}^{(3,4)}_{j}(0)=(0,\pm R, 0 )$,
 ${\bf a}^{(5,6)}_{j}(0)=(0,0,\pm R )$.

The displacement of the center of the $j$-th particle at the time moment $t$
is described by the vector ${\bf r}_{j}(t)$.
The orientation of $j$-th particle at the time moment $t$
is given by the unit quaternion ${\bf \Lambda}_{j}(t)$.
 That quaternion may be represented in the form (see,~e.g.,~\cite{quat})
${\bf \Lambda}_{j}(t)=[\cos(\phi_{j}(t)/2),\sin(\phi_{j}(t)/2) {\bf n }_{j}(t)]$,
where the unit length vector $|{\bf n }_{j}(t)|=1$ describes the
axis, passing through the center of the particle and
$\phi_{j}$ is the angle of rotation around this axis.
We also introduce conjugated quaternion
$\tilde {\bf \Lambda}_{j}(t)=[\cos(\phi_{j}(t)/2),-\sin(\phi_{j}(t)/2) {\bf n }_{j}(t)]$.
Finally, the location of the $k$-th patch of the $j$-th particle
at the time moment $t$ is given by formula
${\bf a}_{j}^{(k)}(t)={\bf r}_{j}(t)+{\bf \Lambda}_{j}(t) \otimes {\bf a}^{(k)}_{j}(0)
\otimes \tilde {\bf \Lambda}_{j}(t)$
where $\otimes$ denotes the quaternion's product.
In our model cores of patchy particles repel each other with
standard   Lennard-Jones potential smoothly truncated at the distance $2R$~\cite{tlj}
with the interaction distance  $\sigma=2R$
and interaction strength  $\epsilon_{0}=1$. Patches
of different particles attract each other with
the  Gaussian potential
$U_{G}({\bf a}_{ij}^{(kl)})=U_{p}
\exp \left[- \left( {\bf a}_{ij}^{(kl)}\right)^{2}/2W^{2}\right]$,
where ${\bf a}_{ij}^{(kl)}={\bf a}_{i}^{(k)}-{\bf a}_{j}^{(l)}$
is a vector connecting a patch $l$ of particle $j$
and a patch $k$ of particle $i$, $W=0.2$ is the half-width of the interaction
and $U_{p}$  is the strength of the interaction.
Knowing the  set of all displacements and orientations of particles
$\{{\bf r}_{j},{\bf \Lambda }_{j} \}$
we can compute a set of total forces
and torques $\{{\bf F}_{j},{\bf M }_{j} \}$ acting on every particles.
We solve  the following  set of kinematic equations
numerically using the velocity Verlet
algorithm
$$
\left\{
\begin{array}{l}
 \dot {\bf v}_{j} (t) = {\bf  F}_{j}(\{ {\bf  r}_{j},{\bf \Lambda}_{j}\})/m\\
\dot { \bf \omega }_{j}(t)= {\bf  M}_{j}(\{ {\bf r}_{j},{\bf \Lambda}_{j}\})/I\\
\end{array}
\right.,
\left\{
\begin{array}{l}
  \dot { \bf r }_{j}(t)  =
{\bf v}_{j}(t)\\
\dot {\bf  \Lambda }_{j}( t)=\frac{1}{2}
{\bf  \omega}_{j}(t) \otimes  {\bf \Lambda }_{j}(t)
\end{array}
\right.,
$$
where ${\bf v}_{j}$ is the velocity of the $j$-th particle and
${\bf  \omega}_{j}$ is its angular velocity,
$\dot {\bf v}_{j}$ and $\dot { \bf \omega }_{j}$ are linear
and angular accelerations, respectively.
We use normalized units: the radius of particles  $R=1$,
the mass is $m=1$, the moment of inertia of a solid sphere
$I=\frac{2}{5}mR^{2}=0.4$.
To simulate the interaction with solvent
we add  Langevin terms $-\gamma {\bf v}_{j}(t)+\xi_{j}(t)$
and $-\frac{4}{3}\gamma R^{2} {\bf \omega}_{j}(t)+\zeta_{j}(t)$
to forces and torques, respectively, where $\gamma=6 \pi \eta R$
is the friction coefficient for solvent viscosity $\eta$,
$ \xi_{j}(t)$ and $ \xi_{j}(t)$ are thermal noise terms
with delta-correlated components
$\left<  \xi_{i}^{\alpha}(t)\xi_{j}^{\beta}(t') \right>=
2\gamma  k_{\rm B}T \delta_{i,j}\delta_{\alpha,\beta}\delta_{t,t'} $,
$\left<  \zeta_{i}^{\alpha}(t)\zeta_{j}^{\beta}(t') \right>=
\frac{8}{3}\gamma R^{2} k_{\rm B}T \delta_{i,j}\delta_{\alpha,\beta}\delta_{t,t'} $.
In our simulations, $k_{\rm B}T=1$
and  $\gamma=10$, therefore for times
$t \gg 1/\gamma=0.1$ the dynamics of a particle is Brownian.
We simulate the system of $N=10^3$ spherical particles of radius
$R = 1$  in a
cubic cell of size $L=48$ with periodic boundary conditions.
The volume fraction is $\phi=4/3 \pi R^{3} N/L^{3}\simeq 0.04$.
We can tune the phase state of this system by varying the strength $U_{p}$ of interaction potential between patches.

{\it Structural properties}~~To define the local structural properties of the system we use the bond order parameter method \cite{stein}, which has been widely used in the context of condensed matter physics \cite{stein}, hard sphere systems~\cite{hs, hs_prb}, complex plasmas~\cite{nat_cp,pu,mugrava}, colloidal suspensions \cite{colloids}, granular media etc. Within this method the rotational invariants  of rank $l$ of both second $q_l(i)$ and third $w_l(i)$ order are calculated for each particle $i$ in the system from the vectors (bonds) connecting its center with the centers of its $N_{\rm nn}(i)$ nearest neighboring particles
\be
q_l(i) = \left ( {4 \pi \over (2l+1)} \sum_{m=-l}^{m=l} \vert~q_{lm}(i)\vert^{2}\right )^{1/2}
\ee
\be
w_l(i) = \hspace{-0.8cm} \sum\limits_{\bmx {cc} _{m_1,m_2,m_3} \\_{ m_1+m_2+m_3=0} \emx} \hspace{-0.8cm} \left [ \bmx {ccc} l&l&l \\
m_1&m_2&m_3 \emx \right] q_{lm_1}(i) q_{lm_2}(i) q_{lm_3}(i),
\label{wig}
\ee
\noindent
where $q_{lm}(i) = N_{\rm nn}(i)^{-1} \sum_{j=1}^{N_{\rm nn}(i)} Y_{lm}({\bf r}_{ij} )$, $Y_{lm}$ are the spherical harmonics and
${\bf r}_{ij} = {\bf r}_i - {\bf r}_j$ are vectors connecting centers of particles $i$ and $j$. We note, that the bond order parameters $w_l \propto q_l^3 $, so, in general, these parameters are much more sensitive to the local orientational order in comparison with $q_l$. Here, to define the structural properties of the patched particles,  we calculate the rotational invariants $q_l$, $w_l$ for each particle using the fixed number $N_{\rm nn}$ of the nearest neighbors (NN):  $N_{\rm nn} = 4, 6$  for the first shell of 4pch and 6pch systems; second shell of both systems has $N_{\rm nn} = 12$. The first shells of the ideal 4pch and 6pch  systems have cubic diamond (CD) and simple cubic (SC) lattices, respectively. The second shell has face centered cubic (FCC) lattice for both types of ideal patchy systems. We note, that formation of hexagonal diamond  (HD) with hexagonal close packing (HCP) second shell
is also possible for the patched system, at least kinetically.

\begin{figure}
\includegraphics[width=8.2cm]{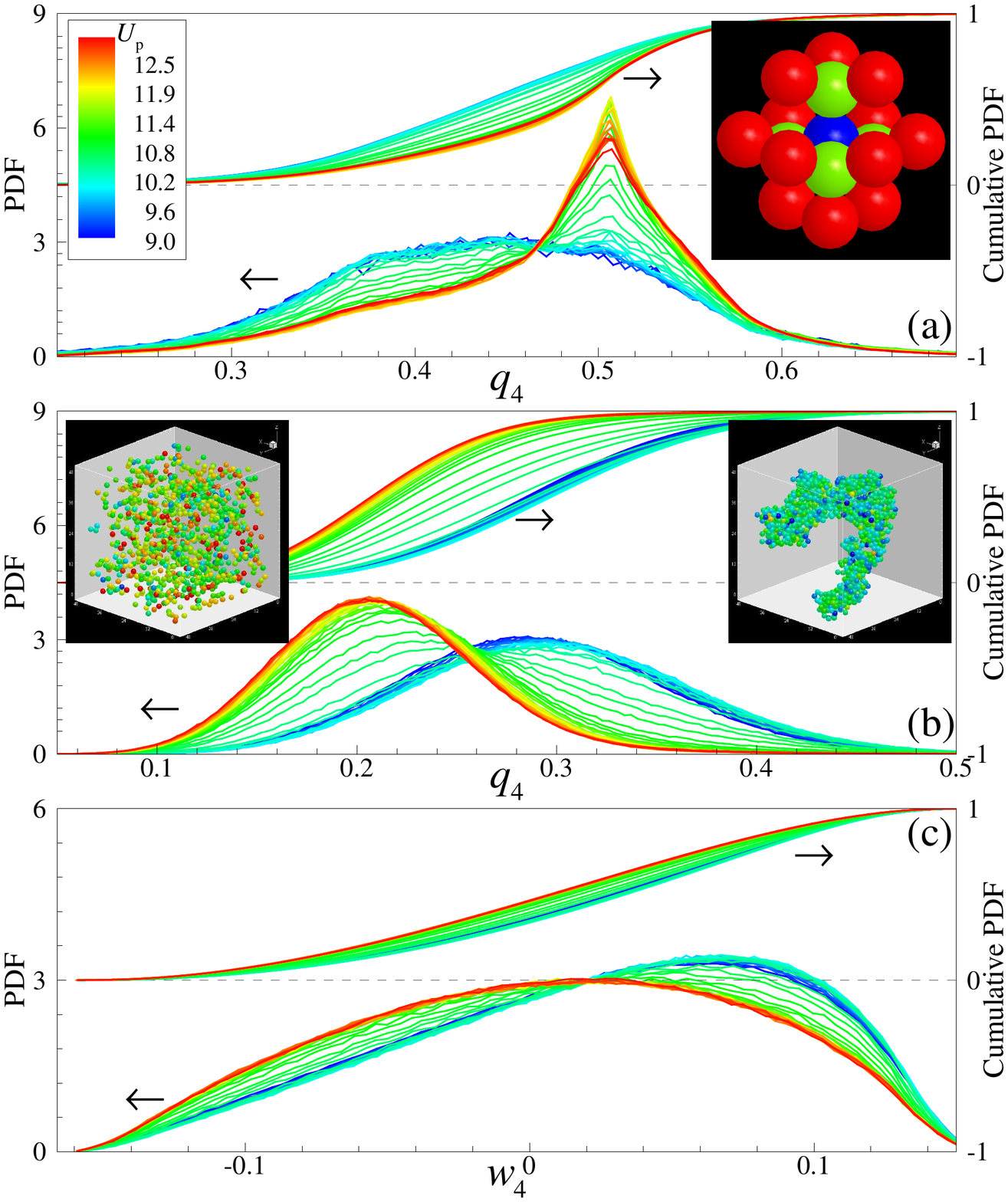}
\caption{(Color online) Patchy system 4pch. Probability distribution functions (PDFs) versus $q_4$ (a,b) and $w_4$ (c) at different strengths $U_p$ of the interaction potential for the first (a) and second (b,c) shells. Cumulative distributions of the PDFs are also plotted to quantify the liquid-solid transition in the system. The curves are color-coded by $U_p$ value. Inset (a) shows first (green) and second (red color) shells of the perfect crystalline particle.
Insets (b) show initial gaseous (b, left) and final gel-like (b, right) particle distribution over space. Particles are color-coded by $q_6$ value.} \label{cl1}
\end{figure}

\begin{figure}
\includegraphics[width=8.2cm]{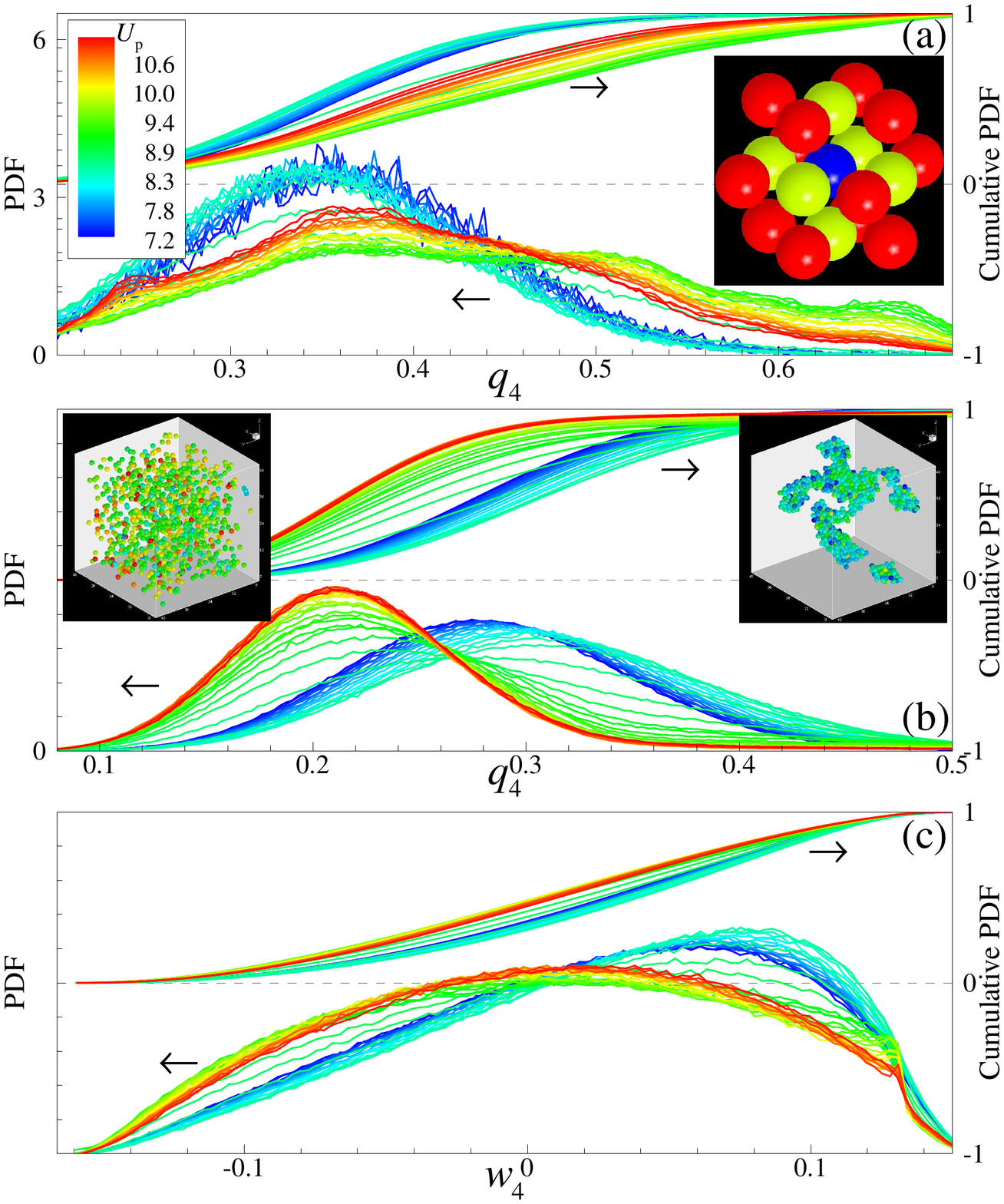}
\caption{(Color online). Patchy system 6pch. Probability distribution functions (PDFs) versus $q_4$ (a,b) and $w_4$ (c)  at different strengths $U_p$ of the interaction potential for the first (a) and second (b,c) shells. Cumulative distributions of the PDFs are also plotted to quantify the liquid-solid transition in the system. The curves are color-coded by $U_p$ value. Inset (a) shows first (green) and second (red color) shells of the perfect  crystalline
particle. Insets (b) show initial gaseous (b, left) and final gel-like (b, right) particle distribution over space. Particles are color-coded by $q_6$ value.} \label{cl2}
\end{figure}

The values of the different rotational invariants $q_l$ and $w_l$ for the perfect patchy crystals (for both first and second shells) are shown in Table 1.
A particle whose coordinates in the 4-dimensional space $(q_4,~ q_6,~ w_4,~ w_6)$ are sufficiently close to those of the ideal lattice is counted as solid-like particle. By calculating the bond order parameters for the second shell it is easy to identify the disordered (liquid-like) phase as well.

\begin{table}[!ht]
  \centering
  \caption{Rotational invariants for the perfect patchy crystals}\label{t:t1}
 \begin{tabular}{ll|cccc}
\hline\hline
system & structure & \quad $q_{4}$ & \quad $q_{6}$ & \quad $w_{4}$ & \quad $ w_{6}$ \\ \hline
4pch & CD, HD (1st shell, 4NN) & 0.509 & 0.628 & -0.159 & -0.013 \\
6pch & SC (1st shell, 6 NN) & 0.76 & 0.35 & 0.159 & 0.013 \\
4,6pch & CD, SC (2nd shell, FCC) & 0.191 & 0.575 & -0.159 & -0.013 \\
4pch & HD (2nd shell, HCP) & 0.097 & 0.485 & 0.134 & -0.012 \\
\hline\hline
\end{tabular}
\end{table}

\begin{figure}
\includegraphics[width=8.2cm]{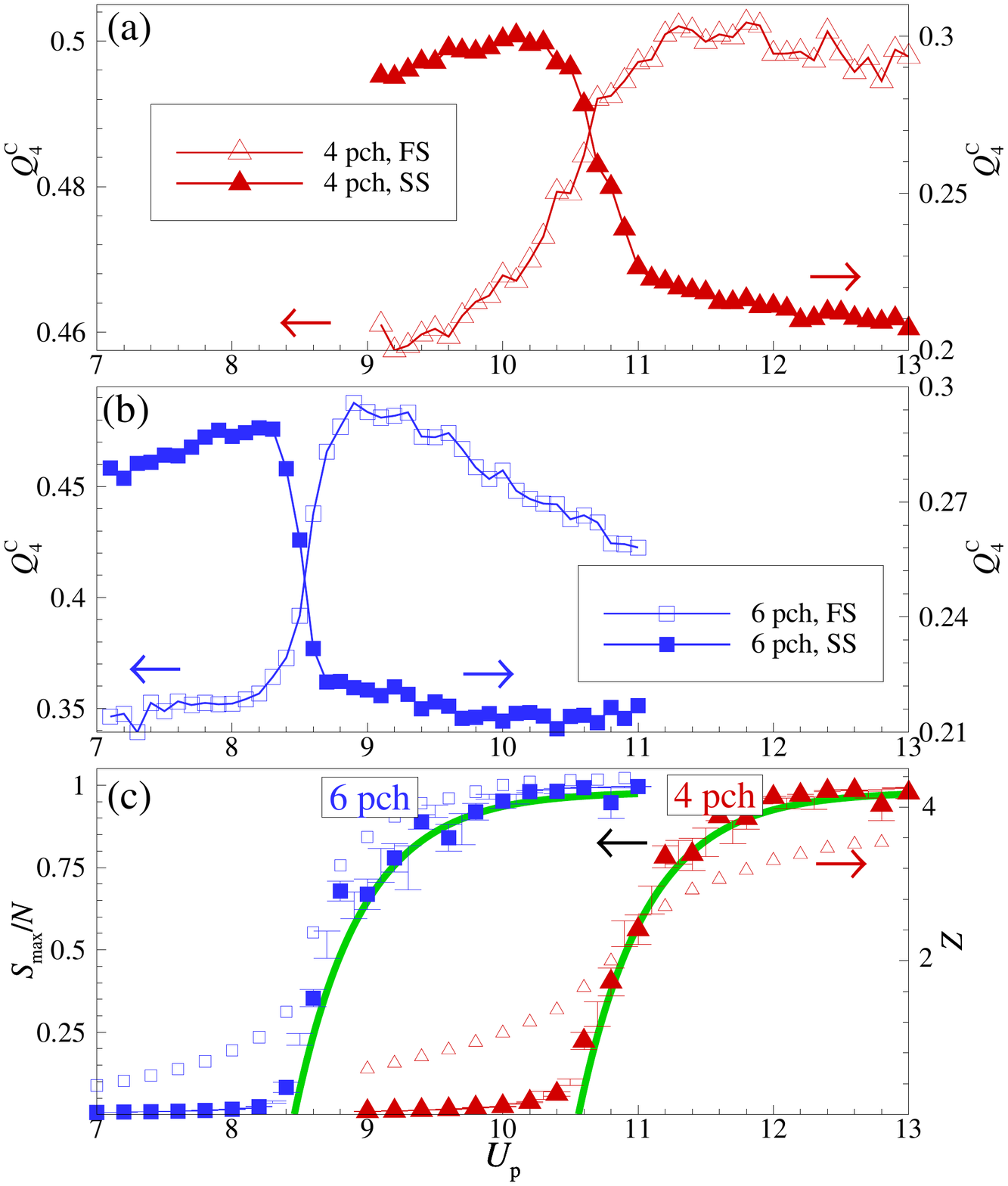}
\caption{(Color online)
(a), (b) The order parameters, characterizing the crystallization of
the system with 4pch (a,c) and 6pch patches (b,c) and associated
with the cumulative PDFs versus $q_4$ value are plotted as a function of the particle interaction potential strength
$U_p$ for both first (FS, open triangles and squares) and second (SS, filled triangles and squares) shells. Panel (c)
shows the normalized maximal cluster size $S_{\rm max}/N$ (solid squares and triangles)  and average coordination number $Z$ (open squares  and triangles)  as functions of $U_p$.
Solid green lines correspond to analytical fitting discussed in text.}
\label{cl}
\end{figure}

Figure~\ref{cl1} and \ref{cl2} show the probability distribution functions (PDFs) of bond order parameters $q_{4}$ (a,b) and $w_{4}$ (c) at different
strengths $U_{p}$ of the interaction potential for the first (a) and second (b,c) shells. Figure~\ref{cl1} and \ref{cl2} correspond to the systems with
4 and 6 patches, respectively. The MD simulations cover both uncorrelated (gas-like) and strongly coupled (liquid-like) phases. The PDFs show
how the structural properties of the ensemble of patched particles vary with increase of interaction strength $U_{p}$.  At low $U_{p}$ (blue lines)
both 4 and 6 patched systems are completely dispersed, and the PDF plots correspond to the isotropic distribution (of uncorrelated system).
Increase of $U_{p}$ results in formation of aggregates of patchy particles; in the final state, nearly all particles consolidate into a few big clusters. \par
Review of the plots for the first shell reveal a remarkable and counterintuitive result. There is a strong evidence of local tetrahedral ordering for 4pch particles, represented by a spike at $q_{4}=0.5$, but very weak order for particles with cubic symmetry. This goes exactly contrary to the known crystallization properties of these systems: as we have discussed earlier, 4pch particles are notoriously hard to crystallize, as opposed to 6pch.
The second shell PDF of $q_{4}$ shows a pronounced but wide peak centered around $q_{4}=0.2$, which is very close to the value expected for ideal FCC lattice. Note that the second shell of both SC and CD lattices has FCC symmetry. It is therefore possible that this is peak is a precursor of the future crystalline order. In order to confirm this correlation, additional studies of a broader class of systems will be needed. If that is the case, it will imply that 4pch system prefers CD structure over HD, at least kinetically. Note that $w_{4}$ does not show any clear signature of FCC order, but that may be due to the fact that it is a higher-order invariant than $q_{4}$. In general both $q_{4} $ and $w_{4}$ PDFs look remarkably similar for both systems, which may originate from the similarity between the second shell structure of SC and CD, but may also be a generic property of amorphous aggregates of patchy particles. It noteworthy that 6pch exhibits very sharp but weak peak at $ w_{4}\approx 0.13$. This value is close to the one for HCP order, but its
actual origin is unclear at the moment. The HCP order is inconsistent with the cubic arrangement of the patches, and $w_{4}$ value alone does not allow
for an unambiguous interpretation.

The cumulative distributions $C_q(q_4)$ and $C_w(w_4)$ are also presented in Figures~\ref{cl1} and \ref{cl2}. One can use the half-height positions for these curves as order parameters that characterize the local orientational order. For instance, $Q_4^{\rm C}$ is defined based on
cumulative distribution of $C_q(q_4)$ as $\int_{-\infty }^{Q_4^{\rm C}} C_q (q_4) dq_4 \equiv  1/2$.  Its value for both first and second shells are plotted in Figure~\ref{cl} as a function of $U_{p}$ for both patchy systems. Note that the completely isotropic distribution corresponds to non-zero values of  $Q_4^{\rm C}$ due to a finite number of particles in each shell. The deviation from that values characterizes the degree of orientational correlations in the system. We observe a clear crossover from isotropic gas phase to strongly correlated liquid at high values of $U_{p}$.

In order to understand the reason for the striking difference in the first-shell orientational order between the two systems, we will now
consider their topological properties. Two patches are defined to be bound if they are closer than $2W$ from each other (where $W=0.2$ is the width
parameter of the Gaussian potential). This allows us to construct clusters of connected particles and compute the fraction of particles that belong to
the maximum cluster, $S_{\rm max}/N$. In addition, we determine the mean coordination number $Z$,  i.e. the average number of
neighbors to which a particle is connected. Both properties are presented in Figure~\ref{cl} together with cumulative bond orientation parameter $Q_4^{\rm C}$ discussed above. Naturally, the signature of aggregation appears simultaneously on all the plots. Remarkably, the coordination number in both systems exhibits saturation at value $Z\approx 4$, despite the particle having dramatically different design and number of patches. This observation is the key to understanding the difference in first shell ordering. Indeed, this value of $Z$ means that 4pch particles in random aggregate have almost the same number of bonds as in ideal diamond crystal, either cubic or hexagonal. Hence, the first shell exhibits strong tetrahedral ordering. On the other hand, at $Z=4$ the 6pch system is far from connectivity of the corresponding simple cubic lattice, and the orientational order is far less pronounced.

The dependence of the largest cluster size on $U_{p}$ can be also related to the value of $Z$. The concentration of free  particles depends exponentially on the chemical potential of those belonging to the aggregate, $\mu =ZU_{p}/2+const$. The size of the biggest cluster can be therefore estimated
by subtracting the gas-like fraction  of the system from the total: $S_{\max }/N=1-\exp \left( -Z\left( U_{p}-U_{0}\right) /2\right) $.
This  formula with Z=4 well describes the numerical results, as shown in Fig. 3(c).
 The relative shift of the two plots can be attributed to an additional entropy of the 6pch liquid. This entropy,
$\Delta S\approx 4k_{\rm B}$ per particle reflects a much  larger number of ways
in which 6-patch system can be arranged into  into $Z=4$ network.

What determines the coordination number of the amorphous aggregate? In the
limit of infinitely short range of interactions between patches, and \
hard-core interparticle repulsion, the bond orientation and relative
positions of the two bound particles would be completely restricted. The
only remaining degree of freedom would be rotation about the direction of
the bond. This means that each perfect bond freezes $5$ translational and
orientational degrees of freedom (or $2d-1$ in $d$ dimensions). Since the
total number of degrees of freedom for $N$ particles is $d\left( d+1\right)
N/2$, the system becomes completely rigid and incapable of creating new
bonds for coordination number $ Z_{\rm rigid}=d\left( d+1\right)/(2d-1)=2.4$ for $d$ = 3.
This result represents conceptually important but somewhat unrealistic case
from both experimental and computational points of view. Both interactions
between patches and interparticle repulsions have finite range which means
that bonds are not infinitely rigid. Interestingly, out of $5$ degrees of
freedom that a perfectly rigid bond would suppress, not all are equally
affected in the case of finite interaction range.  If $\lambda _{1}$ is the
typical range of intra-patch attraction, and $\lambda _{2}$ is that of
interparticle repulsion, a single bond  results in an angular confinement of
a particle within solid angle  $\delta \Omega \simeq 4\pi \left( \lambda
_{1}+\lambda _{2}\right) /R$. This means that each of the  angular
coordinates is constrained much weaker in relative terms, than the
translational degrees of freedom  of the bound patches, $ \delta \theta \simeq \sqrt{\frac{\left( \lambda _{1}+\lambda _{2}\right) }{R}}\gg \frac{\lambda _{1}}{R}$ If we now repeat the above counting argument by only assuming each bond to
suppress $d$ translational degrees of freedom, we obtain a new estimate of
the coordination number of the amorphous aggregate $Z^{\ast } = d+1=4$ for $d$ = 3.
This value of $Z$ is indeed consistent with our results, and also  plays
a prominent role in a other important problems. For instance, it corresponds
to  isostatic packing of objects with infinite friction coefficient.

To summarize, the random  aggregation naturally results in a liquid with
coordination number close to  $Z^{\ast }=4$. This is close to maximum
connectivity of 4pch particles, but is substantially below that for 6pch
system. As a result, the bond orientational order is much more pronounced in
the first shell of particles with tetrahedral symmetry than for those with
cubic ones.  Paradoxically, this also means much lower driving force towards
crystallization in the tetrahedral case. This complements the well known
problem of degeneracy of the ground state of the system: not only is there
a competition between  energetically very similar cubic and hexagonal diamond,
but both of them have very little advantage in connectivity  over generic
aggregate with $Z\approx Z^{\ast }=4$. The 6pch system on the
other hand has a clear energetic incentive to     form a well coordinated SC
\ crystal. Altogether this explains both why \ the diamond is so hard to
self-assemble and also the anomalous orientational order  that we report
for 4pch liquid. In addition, our analysis of the second shell organization
reveal signatures consistent with FCC ordering in both system, which is
indeed expected in the corresponding  crystals,  SC for 6pch and CD for
4pch.  The use of RI-based analysis developed in this work  on a broader
range of patchy systems will be a useful tool to characterize seemingly
structureless aggregates, and capture early precursors of order.

Research is supported by European Research Council under FP7 IRSES
Marie-Curie grants PIRSES-GA-2010-269139 and PIRSES-GA-2010-269181.
BAK was  supported partially by  the Russian Foundation for Basic Research, Project
no. 13-02-00913. Research carried out in part at the Center for Functional
Nanomaterials, Brookhaven National Laboratory, which is supported by the U.S.
Department of Energy, Office of Basic Energy Sciences, under Contract No. DE-AC02-98CH10886.


\end{document}